# Proper curvature symmetry in non-static cylindrically symmetric Lorentzian manifolds

Ghulam Shabbir and M. Ramzan

Faculty of Engineering Sciences, GIK Institute of Engineering
Sciences and Technology, Topi, Swabi, Khyber Pukhtoonkhwa, Pakistan.

Email: shabbir@giki.edu.pk

**Abstract**

We considered the most general form of non-static cylindrically symmetric space-times for studying proper curvature symmetry by using the rank of the $6\times 6$ Riemann matrix and direct integration techniques. Studying proper curvature symmetry in each case of the above space-times it is shown that when the above space-times admit proper curvature symmetry, they form an infinite dimensional vector space.

**PACS numbers**: 04.20.-q, 04.20.Jb

**Key words**: Rank of the $6\times 6$ Rieman matrix; Proper curvature symmetry; Direct integration techinque.

## 1. INTRODUCTION

The aim of this paper is to discuss the proper curvature symmetry of non-static cylindrically symmetric space-times. Curvature symmetry which preserves the curvature structure of a space-time carries significant information and plays an important role in Einstein's theory of general relativity. It is therefore important to study curvature symmetry. Different approaches [1-24] were adopted to study curvature symmetries. Here an aproach, which is given in [7], is adopted to study proper curvature symmetry in non-static cylindrically symmetric space-times by using the rank of the $6\times 6$ Rieman matrix and direct integration techinques. Throughout $M$ represents a four dimensional, connected, Hausdorff space-time manifold with Lorentz metric $g$ of signature (-, +, +, +). The curvature tensor associated with $g_{ab}$, through the Levi-Civita connection, is denoted in component form by $R^a{}_{bcd}$, and the Ricci tensor components are $R_{ab} = R^c{}_{acb}$. The usual covariant, partial and Lie derivatives are denoted by a semicolon, a comma and the symbol $L$, respectively. Round and square brackets denote the usual symmetrization and skew-symmetrization, respectively. Here, $M$ is assumed non-flat in the sense that the curvature tensor does not vanish over any non-empty open subset of $M$.

The covariant derivative of any vector field $X$ on $M$ can be decomposed as

$$X_{a;b} = \frac{1}{2}h_{ab} + G_{ab} \tag{1}$$

where $h_{ab}(=h_{ba}) = L_X g_{ab}$ is a symmetric and $G_{ab}(=-G_{ba})$ is a skew symmetric tensor on $M$. If $h_{ab;c} = 0$, $X$ is said to be *affine* and further satisfies $h_{ab} = 2cg_{ab}, c \in R$ then $X$ is said to be *homothetic* (and *Killing* if $c = 0$). The vector field $X$ is said to be proper affine if it is not homothetic vector field and also $X$ is said to be proper homothetic vector field if it is not Killing vector field.

A vector field $X$ on $M$ is called a curvature symmetry or curvature collineation (CC) if it satisfies [18]

$$L_X R^a{}_{bcd} = 0 \tag{2}$$

or equivalently,

$$R^a{}_{bcd;e} X^e + R^a{}_{ecd} X^e{}_{;b} + R^a{}_{bed} X^e{}_{;c} + R^a{}_{bce} X^e{}_{;d} - R^e{}_{bcd} X^a{}_{;e} = 0.$$

The vector field $X$ is said to be proper curvature symmetry or proper curvature collineation (CC) if it is not affine [7] on $M$.

## 2. CLASSIFICATION OF THE RIEMANN TENSORS

In this section we will classify the Riemann tensor in terms of its rank and bivector decomposition.

The rank of the Riemann tensor is the rank of the $6 \times 6$ symmetric matrix derived in a well known way [7]. The rank of the Riemann tensor at $p$ is the rank of the linear map $\eta$ which maps the vector space of all bivectors $G$ at $p$ to itself and is defined by $\eta : G^{ab} \to R^{ab}{}_{cd} G^{cd}$. Define also the subspace $N_p$ of the tangent space $T_p M$ consisting of those members $k$ of $T_p M$ which satisfy the relation

$$R_{abcd} k^d = 0. \tag{3}$$

Then the Riemann tensor at $p$ satisfies exactly one of the following algebraic conditions [7].

**Class B**

The rank is 2 and the range of $\eta$ is spanned by the dual pair of non-null simple bivectors and $\dim N_p = 0$. The Riemann tensor at $p$ takes the form

$$R_{abcd} = \alpha\, G_{ab} G_{cd} + \beta\, \overset{*}{G}_{ab} \overset{*}{G}_{cd}, \tag{4}$$

where $G$ and its dual $\overset{*}{G}$ are the (unique up to scaling) simple non-null spacelike and timelike bivectors in the range of $\eta$, respectively and $\alpha, \beta \in R$.

**Class C**

The rank is 2 or 3 and there exists a unique (up to scaling) solution say, $k$ of (18) (and so $\dim N_p = 1$). The Riemann tensor at $p$ takes the form

$$R_{abcd} = \sum_{i,j=1}^{3} \alpha_{ij}\, G^{i}{}_{ab}\, G^{j}{}_{cd}, \tag{5}$$

where $\alpha_{ij} \in R$ for all $i, j$ and $G^{i}{}_{ab} k^{b} = 0$ for each of the bivectors $G^{i}$ which span the range of $\eta$.

**Class D**

Here the rank of the curvature matrix is 1. The range of the map $\eta$ is spanned by a single bivector $G$, say, which has to be simple because the symmetry of Riemann tensor $R_{a[bcd]} = 0$ means $G_{a[b} G_{cd]} = 0$. Then it follows from a standard result that $G$ is simple. The curvature tensor admits exactly two independent solutions $k, u$ of equation (3) so that $\dim N_p = 2$. The Riemann tensor at $p$ takes the form

$$R_{abcd} = \alpha\, G_{ab} G_{cd}, \tag{6}$$

where $\alpha \in R$ and $G$ is simple bivector with blade orthogonal to $k$ and $u$.

**Class O**

The rank of the curvature matrix is 0 (so that $R_{abcd} = 0$) and $\dim N_p = 4$.

**Class A**

The Riemann tensor is said to be of class A at $p$ if it is not of class B, C, D or O. Here always $\dim N_p = 0$.

A study of the curvature symmetries or curvature collineations (CCS) for the classes A, B, D, C and O can be found in [7,16].

## 3. MAIN RESULTS

Consider a non static cylindrically symmetric space-time in the usual coordinate system $(t, r, \theta, \phi)$ (labeled by $(x^0, x^1, x^2, x^3)$, respectively) with line element [25]

$$ds^2 = -e^{U(t,r)}dt^2 + dr^2 + e^{V(t,r)}d\theta^2 + e^{W(t,r)}d\phi^2. \tag{7}$$

The Ricci tensor Segre type of the above space-time is $\{1,111\}$ or $\{211\}$ or one of its degeneracies. The above space-time admits two linearly independent Killing vector fields which are

$$\frac{\partial}{\partial \theta}, \frac{\partial}{\partial \phi}. \tag{8}$$

The non-zero independent components of the Riemann tensor are

$$R_{0101} = -\frac{1}{4}\left[U_r^2(t,r) + 2U_{rr}(t,r)\right]e^{U(t,r)} \equiv \alpha_1,$$

$$R_{0202} = -\frac{1}{4}\left[V_t^2(t,r) + 2V_{tt}(t,r) - U_t(t,r)V_t(t,r)\right]e^{V(t,r)} + \frac{1}{4}e^{U(t,r)+V(t,r)}U_r(t,r)V_r(t,r) \equiv \alpha_2,$$

$$R_{0212} = -\frac{1}{4}\left[V_t(t,r)V_r(t,r) + 2V_{tr}(t,r) - U_r(t,r)V_t(t,r)\right]e^{V(t,r)} \equiv \alpha_3,$$

$$R_{0303} = -\frac{1}{4}\left[W_t^2(t,r) + 2W_{tt}(t,r) - U_t(t,r)W_t(t,r)\right]e^{W(t,r)} + \frac{1}{4}e^{U(t,r)+W(t,r)}U_r(t,r)W_r(t,r) \equiv \alpha_4,$$

$$R_{0313} = -\frac{1}{4}\left[W_t(t,r)W_r(t,r) + 2W_{tr}(t,r) - U_r(t,r)W_t(t,r)\right]e^{W(t,r)} \equiv \alpha_5,$$

$$R_{1212} = -\frac{1}{4}\left[V_r^2(t,r) + 2V_{rr}(t,r)\right]e^{V(t,r)} \equiv \alpha_6,$$

$$R_{1313} = -\frac{1}{4}\left[W_r^2(t,r) + 2W_{rr}(t,r)\right]e^{W(t,r)} \equiv \alpha_7,$$

$$R_{2323} = -\frac{1}{4}\left[-V_t(t,r)W_t(t,r) + V_r(t,r)W_r(t,r)\right]e^{V(t,r)+W(t,r)} \equiv \alpha_8.$$

Writing the curvature tensor with components $R_{abcd}$ at $p$ as a $6\times 6$ symmetric matrix

$$R_{abcd} = \begin{pmatrix} \alpha_1 & 0 & 0 & 0 & 0 & 0 \\ 0 & \alpha_2 & 0 & \alpha_3 & 0 & 0 \\ 0 & 0 & \alpha_4 & 0 & \alpha_5 & 0 \\ 0 & \alpha_3 & 0 & \alpha_6 & 0 & 0 \\ 0 & 0 & \alpha_5 & 0 & \alpha_7 & 0 \\ 0 & 0 & 0 & 0 & 0 & \alpha_8 \end{pmatrix}. \tag{9}$$

It is important to remind the reader that we will consider Riemann tensor components as $R^a{}_{bcd}$ for calculating CCS. Here, we are only interested in those cases when the rank of the $6\times 6$ Riemann matrix is less than or equal to three. Since we know from theorem [2,7] that when the rank of the $6\times 6$ Riemann matrix is greater than three there exists no proper curvature symmetries or proper CCS. Thus there exist the following possibilities:

(A1) Rank=3, $U_t(t,r) = 0$, $W_t(t,r) \neq 0$, $U_r(t,r) = 0$, $V_t(t,r) = 0$, $V_r(t,r) \neq 0$, $V_r^2(t,r) + 2V_{rr}(t,r) = 0$, $W_t(t,r)W_r(t,r) + 2W_{tr}(t,r) \neq 0$, $W_t^2(t,r) + 2W_{tt}(t,r) \neq 0$, $W_r(t,r) \neq 0$ and $W_r^2(t,r) + 2W_{rr}(t,r) \neq 0$.

(A2) Rank=3, $U_t(t,r) \neq 0$, $U_r(t,r) = 0$, $V_t(t,r) = 0$, $V_r(t,r) \neq 0$, $V_r^2(t,r) + 2V_{rr}(t,r) = 0$, $W_t(t,r) \neq 0$, $W_r(t,r) \neq 0$, $W_t(t,r)W_r(t,r) + 2W_{tr}(t,r) \neq 0$, $W_t^2(t,r) + 2W_{tt}(t,r) = 0$ and $W_r^2(t,r) + 2W_{rr}(t,r) \neq 0$.

(A3) Rank=3, $U_t(t,r) = 0$, $U_r(t,r) = 0$, $V_t(t,r) \neq 0$, $V_r(t,r) = 0$, $V_t^2(t,r) + 2V_{tt}(t,r) = 0$, $W_t(t,r) \neq 0$, $W_r(t,r) \neq 0$, $W_t(t,r)W_r(t,r) + 2W_{tr}(t,r) = 0$, $W_t^2(t,r) + 2W_{tt}(t,r) \neq 0$ and $W_r^2(t,r) + 2W_{rr}(t,r) \neq 0$.

(A4) Rank=3, $U_t(t,r) = 0$, $U_r(t,r) = 0$, $V_t(t,r) \neq 0$, $V_r(t,r) = 0$, $V_t^2(t,r) + 2V_{tt}(t,r) = 0$, $W_t(t,r) \neq 0$, $W_r(t,r) \neq 0$, $W_t(t,r)W_r(t,r) + 2W_{tr}(t,r) \neq 0$, $W_t^2(t,r) + 2W_{tt}(t,r) = 0$ and $W_r^2(t,r) + 2W_{rr}(t,r) = 0$.

(A5) Rank=3, $U_t(t,r) = 0$, $U_r(t,r) = 0$, $V_t(t,r) \neq 0$, $V_r(t,r) \neq 0$, $V_t^2(t,r) + 2V_{tt}(t,r) = 0$, $V_r^2(t,r) + 2V_{rr}(t,r) = 0$, $V_t(t,r)V_r(t,r) + 2V_{tr}(t,r) = 0$, $W_t(t,r) \neq 0$, $W_r(t,r) \neq 0$, $W_t(t,r)W_r(t,r) + 2W_{tr}(t,r) \neq 0$, $W_t^2(t,r) + 2W_{tt}(t,r) = 0$ and $W_r^2(t,r) + 2W_{rr}(t,r) \neq 0$.

(A6) Rank=3, $U_t(t,r) \neq 0$, $U_r(t,r) = 0$, $V_t(t,r) = 0$, $V_r(t,r) \neq 0$, $V_r^2(t,r) + 2V_{rr}(t,r) = 0$, $W_t(t,r) \neq 0$, $W_r(t,r) \neq 0$, $W_t(t,r)W_r(t,r) + 2W_{tr}(t,r) \neq 0$ and $W_r^2(t,r) + 2W_{rr}(t,r) \neq 0$.

(A7) Rank=3, $U_t(t,r) = 0$, $U_r(t,r) = 0$, $V_t(t,r) \neq 0$, $V_r(t,r) \neq 0$, $V_t^2(t,r) + 2V_{tt}(t,r) = 0$, $V_r^2(t,r) + 2V_{rr}(t,r) = 0$, $V_t(t,r)V_r(t,r) + 2V_{tr}(t,r) = 0$, $W_t(t,r) \neq 0$, $W_r(t,r) \neq 0$, $W_t(t,r)W_r(t,r) + 2W_{tr}(t,r) \neq 0$, $W_t^2(t,r) + 2W_{tt}(t,r) \neq 0$ and $W_r^2(t,r) + 2W_{rr}(t,r) \neq 0$.

(A8) Rank=3, $U_t(t,r) = 0$, $U_r(t,r) = 0$, $V_t(t,r) \neq 0$, $V_r(t,r) = 0$, $V_t^2(t,r) + 2V_{tt}(t,r) = 0$, $W_t(t,r) \neq 0$, $W_r(t,r) \neq 0$, $W_t(t,r)W_r(t,r) + 2W_{tr}(t,r) \neq 0$, $W_t^2(t,r) + 2W_{tt}(t,r) \neq 0$ and $W_r^2(t,r) + 2W_{rr}(t,r) \neq 0$.

(A9) Rank=3, $U_t(t,r) \neq 0$, $U_r(t,r) = 0$, $V_t(t,r) = 0$, $V_r(t,r) \neq 0$, $V_r^2(t,r) + 2V_{rr}(t,r) \neq 0$, $W_t(t,r) \neq 0$, $W_r(t,r) \neq 0$, $W_t(t,r)W_r(t,r) + 2W_{tr}(t,r) = 0$, $W_t^2(t,r) + 2W_{tt}(t,r) = 0$ and $W_r^2(t,r) + 2W_{rr}(t,r) = 0$.

(A10) Rank=3, $U_t(t,r) = 0$, $U_r(t,r) = 0$, $V_t(t,r) = 0$, $V_r(t,r) \neq 0$, $V_r^2(t,r) + 2V_{rr}(t,r) \neq 0$, $W_t(t,r) \neq 0$, $W_r(t,r) \neq 0$, $W_t(t,r)W_r(t,r) + 2W_{tr}(t,r) = 0$, $W_t^2(t,r) + 2W_{tt}(t,r) \neq 0$ and $W_r^2(t,r) + 2W_{rr}(t,r) = 0$.

(A11) Rank=3, $U_t(t,r) = 0$, $U_r(t,r) = 0$, $V_t(t,r) \neq 0$, $V_r(t,r) \neq 0$, $V_t^2(t,r) + 2V_{tt}(t,r) = 0$, $V_r^2(t,r) + 2V_{rr}(t,r) \neq 0$, $V_t(t,r)V_r(t,r) + 2V_{tr}(t,r) = 0$, $W_t(t,r) \neq 0$, $W_r(t,r) \neq 0$, $W_t(t,r)W_r(t,r) + 2W_{tr}(t,r) = 0$, $W_t^2(t,r) + 2W_{tt}(t,r) \neq 0$ and $W_r^2(t,r) + 2W_{rr}(t,r) = 0$.

(A12) Rank=3, $U_t(t,r) = 0$, $U_r(t,r) = 0$, $V_t(t,r) \neq 0$, $V_r(t,r) \neq 0$, $V_t^2(t,r) + 2V_{tt}(t,r) = 0$, $V_r^2(t,r) + 2V_{rr}(t,r) \neq 0$, $V_t(t,r)V_r(t,r) + 2V_{tr}(t,r) = 0$, $W_t(t,r) \neq 0$, $W_r(t,r) = 0$ and $W_t^2(t,r) + 2W_{tt}(t,r) \neq 0$.

(A13) Rank=3, $U_t(t,r) = 0$, $U_r(t,r) = 0$, $V_t(t,r) \neq 0$, $V_r(t,r) = 0$, $V_t^2(t,r) + 2V_{tt}(t,r) \neq 0$, $W_t(t,r) \neq 0$, $W_r(t,r) \neq 0$, $W_t(t,r)W_r(t,r) + 2W_{tr}(t,r) = 0$, $W_t^2(t,r) + 2W_{tt}(t,r) = 0$ and $W_r^2(t,r) + 2W_{rr}(t,r) \neq 0$.

(A14) Rank=3, $U_t(t,r) \neq 0$, $U_r(t,r) = 0$, $V_t(t,r) \neq 0$, $V_r(t,r) \neq 0$, $V_t^2(t,r) + 2V_{tt}(t,r) = 0$, $V_r^2(t,r) + 2V_{rr}(t,r) = 0$, $V_t(t,r)V_r(t,r) + 2V_{tr}(t,r) = 0$, $W_t(t,r) = 0$, $W_r(t,r) \neq 0$ and $W_r^2(t,r) + 2W_{rr}(t,r) \neq 0$.

(A15) Rank=3, $U_t(t,r) = 0$, $U_r(t,r) = 0$, $V_t(t,r) \neq 0$, $V_r(t,r) \neq 0$, $V_t^2(t,r) + 2V_{tt}(t,r) \neq 0$, $V_r^2(t,r) + 2V_{rr}(t,r) = 0$, $V_t(t,r)V_r(t,r) + 2V_{tr}(t,r) = 0$, $W_t(t,r) \neq 0$, $W_r(t,r) \neq 0$, $W_t(t,r)W_r(t,r) + 2W_{tr}(t,r) = 0$, $W_t^2(t,r) + 2W_{tt}(t,r) = 0$ and $W_r^2(t,r) + 2W_{rr}(t,r) \neq 0$.

(A16) Rank=3, $U_t(t,r) = 0$, $U_r(t,r) = 0$, $V_t(t,r) \neq 0$, $V_r(t,r) \neq 0$, $V_t^2(t,r) + 2V_{tt}(t,r) \neq 0$, $V_r^2(t,r) + 2V_{rr}(t,r) \neq 0$, $V_t(t,r)V_r(t,r) + 2V_{tr}(t,r) \neq 0$, $W_t(t,r) \neq 0$, $W_r(t,r) = 0$ and $W_t^2(t,r) + 2W_{tt}(t,r) = 0$.

(A17) Rank=3, $U_t(t,r) = 0$, $U_r(t,r) = 0$, $V_t(t,r) \neq 0$, $V_r(t,r) \neq 0$, $V_t^2(t,r) + 2V_{tt}(t,r) \neq 0$, $V_r^2(t,r) + 2V_{rr}(t,r) \neq 0$, $V_t(t,r)V_r(t,r) + 2V_{tr}(t,r) \neq 0$, $W_t(t,r) = 0$, $W_r(t,r) \neq 0$ and $W_r^2(t,r) + 2W_{rr}(t,r) = 0$.

(A18) Rank=3, $U_t(t,r) = 0$, $U_r(t,r) = 0$, $V_t(t,r) \neq 0$, $V_r(t,r) \neq 0$, $V_t^2(t,r) + 2V_{tt}(t,r) = 0$, $V_r^2(t,r) + 2V_{rr}(t,r) = 0$, $V_t(t,r)V_r(t,r) + 2V_{tr}(t,r) \neq 0$, $W_t(t,r) = 0$, $W_r(t,r) \neq 0$ and $W_r^2(t,r) + 2W_{rr}(t,r) = 0$.

(A19) Rank=3, $U_t(t,r) \neq 0$, $U_r(t,r) = 0$, $V_t(t,r) \neq 0$, $V_r(t,r) \neq 0$, $V_r^2(t,r) + 2V_{rr}(t,r) \neq 0$, $V_t(t,r)V_r(t,r) + 2V_{tr}(t,r) = 0$, $W_t(t,r) = 0$, $W_r(t,r) \neq 0$ and $W_r^2(t,r) + 2W_{rr}(t,r) = 0$.

(A20) Rank=3, $U_t(t,r) \neq 0$, $U_r(t,r) = 0$, $V_t(t,r) \neq 0$, $V_r(t,r) \neq 0$, $V_t^2(t,r) + 2V_{tt}(t,r) = 0$, $V_r^2(t,r) + 2V_{rr}(t,r) \neq 0$, $V_t(t,r)V_r(t,r) + 2V_{tr}(t,r) = 0$, $W_t(t,r) = 0$, $W_r(t,r) \neq 0$ and $W_r^2(t,r) + 2W_{rr}(t,r) = 0$.

(A21) Rank=3, $U_t(t,r) = 0$, $U_r(t,r) = 0$, $V_t(t,r) \neq 0$, $V_r(t,r) \neq 0$, $V_t^2(t,r) + 2V_{tt}(t,r) \neq 0$, $V_r^2(t,r) + 2V_{rr}(t,r) \neq 0$, $V_t(t,r)V_r(t,r) + 2V_{tr}(t,r) = 0$, $W_t(t,r) \neq 0$, $W_r(t,r) = 0$ and $W_t^2(t,r) + 2W_{tt}(t,r) = 0$.

(A22) Rank=3, $U_t(t,r) = 0$, $U_r(t,r) \neq 0$, $U_r^2(t,r) + 2U_{rr}(t,r) = 0$, $V_t(t,r) = 0$, $V_r(t,r) \neq 0$, $V_r^2(t,r) + 2V_{rr}(t,r) = 0$, $W_t(t,r) \neq 0$, $W_r(t,r) = 0$ and $W_t^2(t,r) + 2W_{tt}(t,r) = 0$.

(A23) Rank=3, $U_t(t,r) \neq 0$, $U_r(t,r) \neq 0$, $U_r^2(t,r) + 2U_{rr}(t,r) = 0$, $V_t(t,r) = 0$, $V_r(t,r) \neq 0$, $V_r^2(t,r) + 2V_{rr}(t,r) = 0$, $W_t(t,r) \neq 0$, $W_r(t,r) = 0$ and $W_t^2(t,r) + 2W_{tt}(t,r) = 0$.

(A24) Rank=3, $U_t(t,r) \neq 0$, $U_r^2(t,r) + 2U_{rr}(t,r) = 0$, $V_t(t,r) = 0$, $V_r(t,r) \neq 0$, $V_r^2(t,r) + 2V_{rr}(t,r) = 0$, $W_t(t,r) \neq 0$ and $W_r(t,r) = 0$.

(A25) Rank=3, $U_t(t,r) = 0$, $U_r(t,r) = 0$, $V_t(t,r) \neq 0$, $V_r(t,r) \neq 0$, $V_t^2(t,r) + 2V_{tt}(t,r) = 0$, $V_r^2(t,r) + 2V_{rr}(t,r) = 0$, $V_t(t,r)V_r(t,r) + 2V_{tr}(t,r) \neq 0$, $W_t(t,r) \neq 0$, $W_r(t,r) = 0$ and $W_t^2(t,r) + 2W_{tt}(t,r) = 0$.

(A26) Rank=3, $U_t(t,r) = 0$, $U_r(t,r) = 0$, $V_t(t,r) \neq 0$, $V_r(t,r) \neq 0$, $V_t^2(t,r) + 2V_{tt}(t,r) = 0$, $V_r^2(t,r) + 2V_{rr}(t,r) = 0$, $V_t(t,r)V_r(t,r) + 2V_{tr}(t,r) \neq 0$, $W_t(t,r) = 0$, $W_r(t,r) \neq 0$ and $W_r^2(t,r) + 2W_{rr}(t,r) = 0$.

(A27) Rank=3, $U_t(t,r) = 0$, $U_r(t,r) = 0$, $V_t(t,r) \neq 0$, $V_r(t,r) \neq 0$, $V_t^2(t,r) + 2V_{tt}(t,r) = 0$, $V_r^2(t,r) + 2V_{rr}(t,r) = 0$, $V_t(t,r)V_r(t,r) + 2V_{tr}(t,r) \neq 0$, $W_t(t,r) \neq 0$, $W_r(t,r) \neq 0$, $W_t(t,r)W_r(t,r) + 2W_{tr}(t,r) = 0$, $W_t^2(t,r) + 2W_{tt}(t,r) = 0$ and $W_r^2(t,r) + 2W_{rr}(t,r) = 0$.

(A28) Rank=3, $U_t(t,r) \neq 0$, $U_r(t,r) \neq 0$, $U_r^2(t,r) + 2U_{rr}(t,r) = 0$, $V_t(t,r) \neq 0$, $V_r(t,r) = 0$, $V_t^2(t,r) + 2V_{tt}(t,r) = 0$, $W_t(t,r) = 0$, $W_r(t,r) \neq 0$ and $W_r^2(t,r) + 2W_{rr}(t,r) = 0$.

(A29) Rank=3, $U_t(t,r) = 0$, $U_r(t,r) \neq 0$, $U_r^2(t,r) + 2U_{rr}(t,r) = 0$, $V_t(t,r) \neq 0$, $V_r(t,r) = 0$, $V_t^2(t,r) + 2V_{tt}(t,r) \neq 0$, $W_t(t,r) = 0$, $W_r(t,r) \neq 0$ and $W_r^2(t,r) + 2W_{rr}(t,r) = 0$.

(A30) Rank=3, $U_t(t,r) = 0$, $U_r(t,r) \neq 0$, $U_r^2(t,r) + 2U_{rr}(t,r) = 0$, $V_t(t,r) = 0$, $V_r(t,r) \neq 0$, $V_r^2(t,r) + 2V_{rr}(t,r) = 0$, $W_t(t,r) \neq 0$, $W_r(t,r) = 0$ and $W_t^2(t,r) + 2W_{tt}(t,r) \neq 0$.

(B1) Rank=2, $U_t(t,r) \neq 0$, $U_r(t,r) = 0$, $V_t(t,r) \neq 0$, $V_r(t,r) = 0$, $V_t^2(t,r) + 2V_{tt}(t,r) = 0$, $W_t(t,r) = 0$, $W_r(t,r) \neq 0$ and $W_r^2(t,r) + 2W_{rr}(t,r) \neq 0$.

(B2) Rank=2, $U_t(t,r) \neq 0$, $U_r(t,r) = 0$, $V_t(t,r) \neq 0$, $V_r(t,r) = 0$, $W_t(t,r) = 0$, $W_r(t,r) \neq 0$ and $W_r^2(t,r) + 2W_{rr}(t,r) \neq 0$.

(B3) Rank=2, $U_t(t,r) = 0$, $U_r(t,r) = 0$, $V_t(t,r) \neq 0$, $V_r(t,r) = 0$, $V_t^2(t,r) + 2V_{tt}(t,r) \neq 0$, $W_t(t,r) = 0$, $W_r(t,r) \neq 0$ and $W_r^2(t,r) + 2W_{rr}(t,r) \neq 0$.

(C1) Rank=3, $U_t(t,r) \neq 0$, $U_r(t,r) = 0$, $V_t(t,r) = 0$, $V_r(t,r) \neq 0$, $W_t(t,r) = 0$, $W_r(t,r) \neq 0$, $V_r^2(t,r) + 2V_{rr}(t,r) \neq 0$ and $W_r^2(t,r) + 2W_{rr}(t,r) \neq 0$.

(C2) Rank=3, $U_t(t,r) \neq 0$, $U_r(t,r) = 0$, $V_t(t,r) \neq 0$, $V_r(t,r) = 0$, $W_t(t,r) \neq 0$, $W_r(t,r) = 0$, $V_t^2(t,r) + 2V_{tt}(t,r) = 0$ and $W_t^2(t,r) + 2W_{tt}(t,r) \neq 0$.

(C3) Rank=3, $U_t(t,r) \neq 0$, $U_r(t,r) = 0$, $V_t(t,r) \neq 0$, $V_r(t,r) = 0$, $W_t(t,r) \neq 0$, $W_r(t,r) = 0$ and $W_t^2(t,r) + 2W_{tt}(t,r) = 0$.

(C4) Rank=3, $U_t(t,r) = 0$, $U_r(t,r) \neq 0$, $U_r^2(t,r) + 2U_{rr}(t,r) \neq 0$, $V_t(t,r) = 0$, $V_r(t,r) = 0$, $W_t(t,r) \neq 0$, $W_r(t,r) = 0$ and $W_t^2(t,r) + 2W_{tt}(t,r) = 0$.

(C5) Rank=3, $U_t(t,r) = 0$, $U_r(t,r) \neq 0$, $U_r^2(t,r) + 2U_{rr}(t,r) \neq 0$, $V_t(t,r) = 0$, $V_r(t,r) \neq 0$, $V_r^2(t,r) + 2V_{rr}(t,r) \neq 0$, $W_t(t,r) = 0$ and $W_r(t,r) = 0$,

(C6) Rank=3, $U_t(t,r) \neq 0$, $U_r(t,r) \neq 0$, $U_r^2(t,r) + 2U_{rr}(t,r) \neq 0$, $V_t(t,r) = 0$, $V_r(t,r) \neq 0$, $V_r^2(t,r) + 2V_{rr}(t,r) \neq 0$, $W_t(t,r) = 0$ and $W_r(t,r) = 0$,

(C7) Rank=3, $U_t(t,r) \neq 0$, $U_r(t,r) \neq 0$, $U_r^2(t,r) + 2U_{rr}(t,r) \neq 0$, $V_t(t,r) \neq 0$, $V_r(t,r) \neq 0$ $V_r^2(t,r) + 2V_{rr}(t,r) \neq 0$, $W_t(t,r) = 0$ and $W_r(t,r) = 0$.

(C8) Rank=3, $U_t(t,r) = 0$, $U_r(t,r) \neq 0$, $U_r^2(t,r) + 2U_{rr}(t,r) \neq 0$, $V_t(t,r) \neq 0$, $V_r(t,r) \neq 0$ $V_r^2(t,r) + 2V_{rr}(t,r) \neq 0$, $W_t(t,r) = 0$ and $W_r(t,r) = 0$.

(C9) Rank=3, $U_t(t,r) = 0$, $U_r(t,r) \neq 0$, $U_r^2(t,r) + 2U_{rr}(t,r) \neq 0$, $V_t(t,r) \neq 0$, $V_r(t,r) \neq 0$, $V_r^2(t,r) + 2V_{rr}(t,r) = 0$, $W_t(t,r) = 0$ and $W_r(t,r) = 0$.

(C10) Rank=3, $U_t(t,r) = 0$, $U_r(t,r) \neq 0$, $U_r^2(t,r) + 2U_{rr}(t,r) \neq 0$, $V_t(t,r) \neq 0$, $V_r(t,r) \neq 0$, $V_t^2(t,r) + 2V_{tt}(t,r) = 0$, $V_r^2(t,r) + 2V_{rr}(t,r) = 0$, $W_t(t,r) = 0$ and $W_r(t,r) = 0$.

(C11) Rank=3, $U_t(t,r) \neq 0$, $U_r(t,r) \neq 0$, $U_r^2(t,r) + 2U_{rr}(t,r) \neq 0$, $V_t(t,r) \neq 0$, $V_r(t,r) = 0$, $W_t(t,r) = 0$ and $W_r(t,r) = 0$.

(C12) Rank=3, $U_t(t,r) \neq 0$, $U_r(t,r) \neq 0$, $U_r^2(t,r) + 2U_{rr}(t,r) \neq 0$, $V_t(t,r) \neq 0$, $V_r(t,r) = 0$, $V_t^2(t,r) + 2V_{tt}(t,r) = 0$, $W_t(t,r) = 0$ and $W_r(t,r) = 0$.

(C13) Rank=2, $U_t(t,r) = 0$, $U_r(t,r) = 0$, $V_t(t,r) = 0$, $V_r(t,r) \neq 0$, $V_r^2(t,r) + 2V_{rr}(t,r) = 0$, $W_t(t,r) = 0$, $W_r(t,r) \neq 0$ and $W_r^2(t,r) + 2W_{rr}(t,r) \neq 0$.

(C14) Rank=2, $U_t(t,r) = 0$, $U_r(t,r) = 0$, $V_t(t,r) = 0$, $V_r(t,r) \neq 0$, $V_r^2(t,r) + 2V_{rr}(t,r) \neq 0$, $W_t(t,r) = 0$, $W_r(t,r) \neq 0$ and $W_r^2(t,r) + 2W_{rr}(t,r) = 0$.

(C15) Rank=2, $U_t(t,r) = 0$, $U_r(t,r) = 0$, $V_t(t,r) \neq 0$, $V_r(t,r) = 0$, $V_t^2(t,r) + 2V_{tt}(t,r) = 0$, $W_t(t,r) \neq 0$, $W_r(t,r) = 0$ and $W_t^2(t,r) + 2W_{tt}(t,r) \neq 0$.

(C16) Rank=2, $U_t(t,r) = 0$, $U_r(t,r) = 0$, $V_t(t,r) \neq 0$, $V_r(t,r) = 0$, $V_t^2(t,r) + 2V_{tt}(t,r) \neq 0$, $W_t(t,r) \neq 0$, $W_r(t,r) = 0$ and $W_t^2(t,r) + 2W_{tt}(t,r) = 0$.

(C17) Rank=2, $U_t(t,r) \neq 0$, $U_r(t,r) = 0$, $V_t(t,r) = 0$, $V_r(t,r) = 0$, $W_t(t,r) \neq 0$, $W_r(t,r) \neq 0$, $W_t(t,r)W_r(t,r) + 2W_{tr}(t,r) \neq 0$ and $W_r^2(t,r) + 2W_{rr}(t,r) = 0$.

(C18) Rank=2, $U_t(t,r) \neq 0$, $U_r(t,r) \neq 0$, $U_r^2(t,r) + 2U_{rr}(t,r) = 0$, $V_t(t,r) = 0$, $V_r(t,r) = 0$, $W_t(t,r) \neq 0$, $W_r(t,r) \neq 0$, $W_t^2(t,r) + 2W_{tt}(t,r) = 0$ and $W_r^2(t,r) + 2W_{rr}(t,r) = 0$.

(C19) Rank=2, $U_t(t,r) = 0$, $U_r(t,r) \neq 0$, $U_r^2(t,r) + 2U_{rr}(t,r) = 0$, $V_t(t,r) = 0$, $V_r(t,r) = 0$, $W_t(t,r) \neq 0$, $W_r(t,r) \neq 0$, $W_t^2(t,r) + 2W_{tt}(t,r) = 0$ and $W_r^2(t,r) + 2W_{rr}(t,r) = 0$.

(C20) Rank=2, $U_t(t,r) \neq 0$, $U_r(t,r) \neq 0$, $U_r^2(t,r) + 2U_{rr}(t,r) = 0$, $V_t(t,r) = 0$, $V_r(t,r) = 0$, $W_t(t,r) \neq 0$, $W_r(t,r) \neq 0$ and $W_r^2(t,r) + 2W_{rr}(t,r) = 0$.

(C21) Rank=2, $U_t(t,r) = 0$, $U_r(t,r) \neq 0$, $U_r^2(t,r) + 2U_{rr}(t,r) = 0$, $V_t(t,r) = 0$, $V_r(t,r) = 0$, $W_t(t,r) \neq 0$, $W_r(t,r) \neq 0$ and $W_r^2(t,r) + 2W_{rr}(t,r) = 0$.

(C22) Rank=2, $U_t(t,r) = 0$, $U_r(t,r) \neq 0$, $U_r^2(t,r) + 2U_{rr}(t,r) = 0$, $V_t(t,r) = 0$, $V_r(t,r) = 0$, $W_t(t,r) \neq 0$, $W_r(t,r) \neq 0$ and $W_r^2(t,r) + 2W_{rr}(t,r) \neq 0$.

(C23) Rank=2, $U_t(t,r) \neq 0$, $U_r(t,r) \neq 0$, $U_r^2(t,r) + 2U_{rr}(t,r) = 0$, $V_t(t,r) = 0$, $V_r(t,r) = 0$, $W_t(t,r) \neq 0$, $W_r(t,r) \neq 0$ and $W_r^2(t,r) + 2W_{rr}(t,r) \neq 0$.

(C24) Rank=2, $U_t(t,r) \neq 0$, $U_r(t,r) \neq 0$, $U_r^2(t,r) + 2U_{rr}(t,r) = 0$, $V_t(t,r) = 0$, $V_r(t,r) = 0$, $W_t(t,r) \neq 0$, $W_r(t,r) \neq 0$, $W_t^2(t,r) + 2W_{tt}(t,r) = 0$ and $W_r^2(t,r) + 2W_{rr}(t,r) \neq 0$.

(C25) Rank=2, $U_t(t,r) = 0$, $U_r(t,r) = 0$, $V_t(t,r) = 0$, $V_r(t,r) = 0$, $W_t(t,r) \neq 0$, $W_r(t,r) \neq 0$, $W_t(t,r)W_r(t,r) + 2W_{tr}(t,r) \neq 0$, $W_t^2(t,r) + 2W_{tt}(t,r) \neq 0$ and $W_r^2(t,r) + 2W_{rr}(t,r) \neq 0$.

(C26) Rank=2, $U_t(t,r) \neq 0$, $U_r(t,r) = 0$, $V_t(t,r) = 0$, $V_r(t,r) = 0$, $W_t(t,r) \neq 0$, $W_r(t,r) \neq 0$, $W_t(t,r)W_r(t,r) + 2W_{tr}(t,r) \neq 0$ and $W_r^2(t,r) + 2W_{rr}(t,r) \neq 0$.

(C27) Rank=2, $U_t(t,r) \neq 0$, $U_r(t,r) = 0$, $V_t(t,r) = 0$, $V_r(t,r) = 0$, $W_t(t,r) \neq 0$, $W_r(t,r) \neq 0$, $W_t(t,r)W_r(t,r) + 2W_{tr}(t,r) = 0$, $W_t^2(t,r) + 2W_{tt}(t,r) = 0$ and $W_r^2(t,r) + 2W_{rr}(t,r) \neq 0$.

(C28) Rank=2, $U_t(t,r) = 0$, $U_r(t,r) = 0$, $V_t(t,r) = 0$, $V_r(t,r) = 0$, $W_t(t,r) \neq 0$, $W_r(t,r) \neq 0$, $W_t(t,r)W_r(t,r) + 2W_{tr}(t,r) \neq 0$, $W_t^2(t,r) + 2W_{tt}(t,r) = 0$ and $W_r^2(t,r) + 2W_{rr}(t,r) \neq 0$.

(C29) Rank=2, $U_t(t,r) = 0$, $U_r(t,r) = 0$, $V_t(t,r) = 0$, $V_r(t,r) = 0$, $W_t(t,r) \neq 0$, $W_r(t,r) \neq 0$, $W_t(t,r)W_r(t,r) + 2W_{tr}(t,r) = 0$, $W_t^2(t,r) + 2W_{tt}(t,r) \neq 0$ and $W_r^2(t,r) + 2W_{rr}(t,r) \neq 0$.

(C30) Rank=2, $U_t(t,r) = 0$, $U_r(t,r) = 0$, $V_t(t,r) = 0$, $V_r(t,r) = 0$, $W_t(t,r) \neq 0$, $W_r(t,r) \neq 0$, $W_t(t,r)W_r(t,r) + 2W_{tr}(t,r) \neq 0$, $W_t^2(t,r) + 2W_{tt}(t,r) = 0$ and $W_r^2(t,r) + 2W_{rr}(t,r) = 0$.

(C31) Rank=2, $U_t(t,r) \neq 0$, $U_r(t,r) = 0$, $V_t(t,r) = 0$, $V_r(t,r) = 0$, $W_t(t,r) \neq 0$, $W_r(t,r) \neq 0$, $W_t(t,r)W_r(t,r) + 2W_{tr}(t,r) = 0$, $W_t^2(t,r) + W_{tt}(t,r) = 0$ and $W_r^2(t,r) + 2W_{rr}(t,r) \neq 0$.

(C32) Rank=2, $U_t(t,r) \neq 0$, $U_r(t,r) \neq 0$, $U_r^2(t,r) + 2U_{rr}(t,r) \neq 0$, $V_t(t,r) = 0$, $V_r(t,r) = 0$, $W_t(t,r) = 0$, $W_r(t,r) \neq 0$ and $W_r^2(t,r) + 2W_{rr}(t,r) = 0$.

(C33) Rank=2, $U_t(t,r) = 0$, $U_r(t,r) \neq 0$, $U_r^2(t,r) + 2U_{rr}(t,r) \neq 0$, $V_t(t,r) = 0$, $V_r(t,r) = 0$, $W_t(t,r) = 0$, $W_r(t,r) \neq 0$ and $W_r^2(t,r) + 2W_{rr}(t,r) = 0$.

(C34) Rank=2, $U_t(t,r) \neq 0$, $U_r(t,r) = 0$, $V_t(t,r) \neq 0$, $V_r(t,r) \neq 0$, $V_t(t,r)V_r(t,r) + 2V_{tr}(t,r) \neq 0$, $V_r^2(t,r) + 2V_{rr}(t,r) \neq 0$, $W_t(t,r) = 0$ and $W_r(t,r) = 0$.

(C35) Rank=2, $U_t(t,r) = 0$, $U_r(t,r) = 0$, $V_t(t,r) \neq 0$, $V_r(t,r) \neq 0$, $V_t^2(t,r) + 2V_{tt}(t,r) \neq 0$, $V_t(t,r)V_r(t,r) + 2V_{tr}(t,r) \neq 0$, $V_r^2(t,r) + 2V_{rr}(t,r) \neq 0$, $W_t(t,r) = 0$ and $W_r(t,r) = 0$.

(C36) Rank=2, $U_t(t,r) \neq 0$, $U_r(t,r) = 0$, $V_t(t,r) \neq 0$, $V_r(t,r) \neq 0$, $V_t^2(t,r) + 2V_{tt}(t,r) = 0$, $V_t(t,r)V_r(t,r) + 2V_{tr}(t,r) \neq 0$, $V_r^2(t,r) + 2V_{rr}(t,r) \neq 0$, $W_t(t,r) = 0$ and $W_r(t,r) = 0$.

(C37) Rank=2, $U_t(t,r) \neq 0$, $U_r(t,r) = 0$, $V_t(t,r) \neq 0$, $V_r(t,r) \neq 0$, $V_t(t,r)V_r(t,r) + 2V_{tr}(t,r) \neq 0$, $V_r^2(t,r) + 2V_{rr}(t,r) = 0$, $W_t(t,r) = 0$ and $W_r(t,r) = 0$.

(C38) Rank=2, $U_t(t,r) \neq 0$, $U_r(t,r) = 0$, $V_t(t,r) \neq 0$, $V_r(t,r) \neq 0$, $V_t^2(t,r) + 2V_{tt}(t,r) = 0$, $V_t(t,r)V_r(t,r) + 2V_{tr}(t,r) \neq 0$, $V_r^2(t,r) + 2V_{rr}(t,r) = 0$, $W_t(t,r) = 0$ and $W_r(t,r) = 0$.

(C39) Rank=2, $U_t(t,r) = 0$, $U_r(t,r) = 0$, $V_t(t,r) \neq 0$, $V_r(t,r) \neq 0$, $V_t^2(t,r) + 2V_{tt}(t,r) \neq 0$, $V_t(t,r)V_r(t,r) + 2V_{tr}(t,r) = 0$, $V_r^2(t,r) + 2V_{rr}(t,r) \neq 0$, $W_t(t,r) = 0$ and $W_r(t,r) = 0$.

(C40) Rank=2, $U_t(t,r) \neq 0$, $U_r(t,r) = 0$, $V_t(t,r) \neq 0$, $V_r(t,r) \neq 0$, $V_t^2(t,r) + 2V_{tt}(t,r) = 0$, $V_t(t,r)V_r(t,r) + 2V_{tr}(t,r) = 0$, $V_r^2(t,r) + 2V_{rr}(t,r) \neq 0$, $W_t(t,r) = 0$ and $W_r(t,r) = 0$.

(C41) Rank=2, $U_t(t,r) = 0$, $U_r(t,r) = 0$, $V_t(t,r) \neq 0$, $V_r(t,r) \neq 0$, $V_t^2(t,r) + 2V_{tt}(t,r) = 0$, $V_t(t,r)V_r(t,r) + 2V_{tr}(t,r) \neq 0$, $V_r^2(t,r) + 2V_{rr}(t,r) = 0$, $W_t(t,r) = 0$ and $W_r(t,r) = 0$.

(C42) Rank=2, $U_t(t,r) \neq 0$, $U_r(t,r) \neq 0$, $U_r^2(t,r) + 2U_{rr}(t,r) \neq 0$, $V_t(t,r) = 0$, $V_r(t,r) \neq 0$, $V_r^2(t,r) + 2V_{rr}(t,r) = 0$, $W_t(t,r) = 0$ and $W_r(t,r) = 0$.

(C43) Rank=3, $U_t(t,r) = 0$, $U_r(t,r) = 0$, $V_t(t,r) \neq 0$, $V_r(t,r) \neq 0$, $V_t^2(t,r) + 2V_{tt}(t,r) = 0$, $V_t(t,r)V_r(t,r) + 2V_{tr}(t,r) = 0$, $V_r^2(t,r) + 2V_{rr}(t,r) \neq 0$, $W_t(t,r) \neq 0$, $W_r(t,r) \neq 0$, $W_t(t,r)W_r(t,r) + 2W_{tr}(t,r) = 0$, $W_t^2(t,r) + 2W_{tt}(t,r) = 0$ and $W_r^2(t,r) + 2W_{rr}(t,r) \neq 0$.

(C44) Rank=3, $U_t(t,r) = 0$, $U_r(t,r) = 0$, $V_t(t,r) \neq 0$, $V_r(t,r) = 0$, $V_t^2(t,r) + 2V_{tt}(t,r) \neq 0$, $W_t(t,r) \neq 0$, $W_r(t,r) \neq 0$, $W_t(t,r)W_r(t,r) + 2W_{tr}(t,r) = 0$, $W_r^2(t,r) + 2W_{rr}(t,r) = 0$ and $W_t^2(t,r) + 2W_{tt}(t,r) \neq 0$.

(C45) Rank=3, $U_t(t,r) \neq 0$, $U_r(t,r) = 0$, $V_t(t,r) \neq 0$, $V_r(t,r) = 0$, $V_t^2(t,r) + 2V_{tt}(t,r) = 0$, $W_t(t,r) \neq 0$, $W_r(t,r) \neq 0$, $W_t(t,r)W_r(t,r) + 2W_{tr}(t,r) = 0$ and $W_r^2(t,r) + 2W_{rr}(t,r) = 0$.

(C46) Rank=3, $U_t(t,r) \neq 0$, $U_r(t,r) = 0$, $V_t(t,r) \neq 0$, $V_r(t,r) = 0$, $W_t(t,r) \neq 0$, $W_r(t,r) \neq 0$, $W_t(t,r)W_r(t,r) + 2W_{tr}(t,r) = 0$, $W_r^2(t,r) + 2W_{rr}(t,r) = 0$ and $W_t^2(t,r) + 2W_{tt}(t,r) = 0$.

(C47) Rank=3, $U_t(t,r) \neq 0$, $U_r(t,r) = 0$, $V_t(t,r) \neq 0$, $V_r(t,r) = 0$, $V_t^2(t,r) + 2V_{tt}(t,r) = 0$, $W_t(t,r) \neq 0$, $W_r(t,r) \neq 0$, $W_t(t,r)W_r(t,r) + 2W_{tr}(t,r) = 0$, $W_t^2(t,r) + 2W_{tt}(t,r) = 0$ and $W_r^2(t,r) + 2W_{rr}(t,r) = 0$.

(C48) Rank=3, $U_t(t,r) = 0$, $U_r(t,r) = 0$, $V_t(t,r) \neq 0$, $V_r(t,r) \neq 0$, $V_t^2(t,r) + 2V_{tt}(t,r) \neq 0$, $V_t(t,r)V_r(t,r) + 2V_{tr}(t,r) = 0$, $V_r^2(t,r) + 2V_{rr}(t,r) = 0$, $W_t(t,r) \neq 0$, $W_r(t,r) = 0$ and $W_t^2(t,r) + 2W_{tt}(t,r) \neq 0$.

(C49) Rank=3, $U_t(t,r) \neq 0$, $U_r(t,r) = 0$, $V_t(t,r) \neq 0$, $V_r(t,r) \neq 0$, $V_t^2(t,r) + 2V_{tt}(t,r) = 0$, $V_t(t,r)V_r(t,r) + 2V_{tr}(t,r) = 0$, $V_r^2(t,r) + 2V_{rr}(t,r) = 0$, $W_t(t,r) \neq 0$ and $W_r(t,r) = 0$.

(C50) Rank=3, $U_t(t,r) \neq 0$, $U_r(t,r) = 0$, $V_t(t,r) \neq 0$, $V_r(t,r) \neq 0$, $V_t(t,r)V_r(t,r) + 2V_{tr}(t,r) = 0$, $V_r^2(t,r) + 2V_{rr}(t,r) = 0$, $W_t(t,r) \neq 0$, $W_r(t,r) = 0$ and $W_t^2(t,r) + 2W_{tt}(t,r) = 0$.

(C51) Rank=3, $U_t(t,r) \neq 0$, $U_r(t,r) = 0$, $V_t(t,r) \neq 0$, $V_r(t,r) \neq 0$, $V_t^2(t,r) + 2V_{tt}(t,r) = 0$, $V_t(t,r)V_r(t,r) + 2V_{tr}(t,r) = 0$, $V_r^2(t,r) + 2V_{rr}(t,r) = 0$, $W_t(t,r) \neq 0$, $W_r(t,r) = 0$ and $W_t^2(t,r) + 2W_{tt}(t,r) = 0$.

(C52) Rank=3, $U_t(t,r) \neq 0$, $U_r(t,r) = 0$, $V_t(t,r) \neq 0$, $V_r(t,r) \neq 0$, $V_t(t,r)V_r(t,r) + 2V_{tr}(t,r) = 0$, $V_r^2(t,r) + 2V_{rr}(t,r) = 0$, $W_t(t,r) \neq 0$, $W_r(t,r) \neq 0$, $W_t(t,r)W_r(t,r) + 2W_{tr}(t,r) = 0$ and $W_r^2(t,r) + 2W_{rr}(t,r) = 0$.

(C53) Rank=3, $U_t(t,r) = 0$, $U_r(t,r) \neq 0$, $U_r^2(t,r) + 2U_{rr}(t,r) = 0$, $V_t(t,r) = 0$, $V_r(t,r) \neq 0$, $V_r^2(t,r) + 2V_{rr}(t,r) = 0$, $W_t(t,r) = 0$, $W_r(t,r) \neq 0$ and $W_r^2(t,r) + 2W_{rr}(t,r) = 0$.

(C54) Rank=2, $U_t(t,r) = 0$, $U_r(t,r) = 0$, $V_t(t,r) \neq 0$, $V_r(t,r) = 0$, $V_t^2(t,r) + 2V_{tt}(t,r) = 0$, $W_t(t,r) \neq 0$, $W_r(t,r) \neq 0$, $W_t(t,r)W_r(t,r) + 2W_{tr}(t,r) = 0$, $W_t^2(t,r) + 2W_{tt}(t,r) = 0$ and $W_r^2(t,r) + 2W_{rr}(t,r) \neq 0$.

(C55) Rank=2, $U_t(t,r) = 0$, $U_r(t,r) = 0$, $V_t(t,r) \neq 0$, $V_r(t,r) \neq 0$, $V_t^2(t,r) + 2V_{tt}(t,r) = 0$, $V_t(t,r)V_r(t,r) + 2V_{tr}(t,r) = 0$, $V_r^2(t,r) + 2V_{rr}(t,r) = 0$, $W_t(t,r) = 0$, $W_r(t,r) \neq 0$ and $W_r^2(t,r) + 2W_{rr}(t,r) \neq 0$.

(C56) Rank=2, $U_t(t,r) = 0$, $U_r(t,r) = 0$, $V_t(t,r) \neq 0$, $V_r(t,r) \neq 0$, $V_t^2(t,r) + 2V_{tt}(t,r) = 0$, $V_t(t,r)V_r(t,r) + 2V_{tr}(t,r) = 0$, $V_r^2(t,r) + 2V_{rr}(t,r) = 0$, $W_t(t,r) \neq 0$, $W_r(t,r) \neq 0$, $W_t(t,r)W_r(t,r) + 2W_{tr}(t,r) = 0$, $W_t^2(t,r) + 2W_{tt}(t,r) = 0$ and $W_r^2(t,r) + 2W_{rr}(t,r) \neq 0$.

(C57) Rank=2, $U_t(t,r) = 0$, $U_r(t,r) = 0$, $V_t(t,r) \neq 0$, $V_r(t,r) \neq 0$, $V_t^2(t,r) + 2V_{tt}(t,r) = 0$, $V_t(t,r)V_r(t,r) + 2V_{tr}(t,r) = 0$, $V_r^2(t,r) + 2V_{rr}(t,r) \neq 0$, $W_t(t,r) \neq 0$, $W_r(t,r) = 0$ and $W_t^2(t,r) + 2W_{tt}(t,r) = 0$.

(C58) Rank=2, $U_t(t,r) = 0$, $U_r(t,r) = 0$, $V_t(t,r) \neq 0$, $V_r(t,r) \neq 0$, $V_t^2(t,r) + 2V_{tt}(t,r) = 0$, $V_t(t,r)V_r(t,r) + 2V_{tr}(t,r) = 0$, $V_r^2(t,r) + 2V_{rr}(t,r) \neq 0$, $W_t(t,r) = 0$, $W_r(t,r) \neq 0$ and $W_r^2(t,r) + 2W_{rr}(t,r) = 0$.

(C59) Rank=2, $U_t(t,r) = 0$, $U_r(t,r) = 0$, $V_t(t,r) = 0$, $V_r(t,r) \neq 0$, $V_r^2(t,r) + 2V_{rr}(t,r) \neq 0$, $W_t(t,r) \neq 0$, $W_r(t,r) \neq 0$, $W_t(t,r)W_r(t,r) + 2W_{tr}(t,r) = 0$ and $W_r^2(t,r) + 2W_{rr}(t,r) = 0$.

(C60) Rank=2, $U_t(t,r) = 0$, $U_r(t,r) = 0$, $V_t(t,r) \neq 0$, $V_r(t,r) \neq 0$, $V_t^2(t,r) + 2V_{tt}(t,r) \neq 0$, $V_t(t,r)V_r(t,r) + 2V_{tr}(t,r) = 0$, $V_r^2(t,r) + 2V_{rr}(t,r) = 0$, $W_t(t,r) \neq 0$, $W_r(t,r) = 0$ and $W_t^2(t,r) + 2W_{tt}(t,r) \neq 0$.

(C61) Rank=2, $U_t(t,r) = 0$, $U_r(t,r) = 0$, $V_t(t,r) \neq 0$, $V_r(t,r) \neq 0$, $V_t^2(t,r) + 2V_{tt}(t,r) \neq 0$, $V_t(t,r)V_r(t,r) + 2V_{tr}(t,r) = 0$, $V_r^2(t,r) + 2V_{rr}(t,r) = 0$, $W_t(t,r) \neq 0$, $W_r(t,r) = 0$ and $W_t^2(t,r) + 2W_{tt}(t,r) = 0$.

(C62) Rank=2, $U_t(t,r) = 0$, $U_r(t,r) = 0$, $V_t(t,r) \neq 0$, $V_r(t,r) \neq 0$, $V_t^2(t,r) + 2V_{tt}(t,r) = 0$, $V_t(t,r)V_r(t,r) + 2V_{tr}(t,r) = 0$, $V_r^2(t,r) + 2V_{rr}(t,r) = 0$, $W_t(t,r) = 0$, $W_r(t,r) \neq 0$ and $W_r^2(t,r) + 2W_{rr}(t,r) = 0$.

(C63) Rank=2, $U_t(t,r) = 0$, $U_r(t,r) = 0$, $V_t(t,r) \neq 0$, $V_r(t,r) \neq 0$, $V_t^2(t,r) + 2V_{tt}(t,r) \neq 0$, $V_t(t,r)V_r(t,r) + 2V_{tr}(t,r) = 0$, $V_r^2(t,r) + 2V_{rr}(t,r) = 0$, $W_t(t,r) = 0$, $W_r(t,r) \neq 0$ and $W_r^2(t,r) + 2W_{rr}(t,r) = 0$.

(C64) Rank=2, $U_t(t,r) = 0$, $U_r(t,r) = 0$, $V_t(t,r) \neq 0$, $V_r(t,r) = 0$, $V_t^2(t,r) + 2V_{tt}(t,r) \neq 0$, $W_t(t,r) \neq 0$, $W_r(t,r) \neq 0$, $W_t(t,r)W_r(t,r) + 2W_{tr}(t,r) = 0$, $W_t^2(t,r) + 2W_{tt}(t,r) = 0$ and $W_r^2(t,r) + 2W_{rr}(t,r) = 0$.

(D1) Rank=1, $U_t(t,r) = 0$, $U_r(t,r) = 0$, $V_t(t,r) = 0$, $V_r(t,r) = 0$, $W_t(t,r) = 0$, $W_r(t,r) \neq 0$ and $W_r^2(t,r) + 2W_{rr}(t,r) \neq 0$.

(D2) Rank=1, $U_t(t,r) \neq 0$, $U_r(t,r) \neq 0$, $U_r^2(t,r) + 2U_{rr}(t,r) \neq 0$, $V_t(t,r) = 0$, $V_r(t,r) = 0$, $W_t(t,r) = 0$ and $W_r(t,r) = 0$.

(D3) Rank=1, $U_t(t,r) = 0$, $U_r(t,r) = 0$, $V_t(t,r) \neq 0$, $V_r(t,r) = 0$, $W_t(t,r) = 0$, $W_r(t,r) = 0$ and $V_t^2(t,r) + 2V_{tt}(t,r) \neq 0$.

(D4) Rank=1, $U_t(t,r) \neq 0$, $U_r(t,r) = 0$, $V_t(t,r) = 0$, $V_r(t,r) = 0$, $W_t(t,r) = 0$, $W_r(t,r) = 0$ and $V_t^2(t,r) + 2V_{tt}(t,r) = 0$.

(D5) Rank=1, $U_t(t,r) \neq 0$, $U_r(t,r) = 0$, $V_t(t,r) = 0$, $V_r(t,r) = 0$, $W_t(t,r) \neq 0$, $W_r(t,r) = 0$ and $W_t^2(t,r) + 2W_{tt}(t,r) = 0$.

(D6) Rank=1, $U_t(t,r) = 0$, $U_r(t,r) = 0$, $V_t(t,r) = 0$, $V_r(t,r) = 0$, $W_t(t,r) \neq 0$, $W_r(t,r) = 0$ and $W_t^2(t,r) + 2W_{tt}(t,r) \neq 0$.

(D7) Rank=1, $U_t(t,r) = 0$, $U_r(t,r) = 0$, $V_t(t,r) = 0$, $V_r(t,r) \neq 0$, $W_t(t,r) = 0$, $W_r(t,r) = 0$ and $V_r^2(t,r) + 2V_{rr}(t,r) \neq 0$.

(D8) Rank=1, $U_t(t,r) = 0$, $U_r(t,r) \neq 0$, $U_r^2(t,r) + 2U_{rr}(t,r) \neq 0$, $V_t(t,r) = 0$, $V_r(t,r) = 0$, $W_t(t,r) = 0$ and $W_r(t,r) = 0$.

(D9) Rank=1, $U_t(t,r) \neq 0$, $U_r(t,r) = 0$, $V_t(t,r) \neq 0$, $V_r(t,r) \neq 0$, $W_t(t,r) = 0$, $W_r(t,r) = 0$, $V_t(t,r)V_r(t,r) + 2V_{tr}(t,r) = 0$ and $V_r^2(t,r) + 2V_{rr}(t,r) = 0$.

(D10) Rank=1, $U_t(t,r) = 0$, $U_r(t,r) = 0$, $V_t(t,r) \neq 0$, $V_r(t,r) \neq 0$, $W_t(t,r) = 0$, $W_r(t,r) = 0$, $V_t(t,r)V_r(t,r) + 2V_{tr}(t,r) = 0$ $V_r^2(t,r) + 2V_{rr}(t,r) = 0$ and $V_t^2(t,r) + 2V_{tt}(t,r) \neq 0$.

(D11) Rank=1, $U_t(t,r) = 0$, $U_r(t,r) = 0$, $V_t(t,r) = 0$, $V_r(t,r) = 0$, $W_t(t,r) \neq 0$, $W_r(t,r) \neq 0$, $W_t(t,r)W_r(t,r) + 2W_{tr}(t,r) = 0$, $W_r^2(t,r) + 2W_{rr}(t,r) = 0$ and $W_t^2(t,r) + 2W_{tt}(t,r) \neq 0$.

(D12) Rank=1, $U_t(t,r) \neq 0$, $U_r(t,r) = 0$, $V_t(t,r) = 0$, $V_r(t,r) = 0$, $W_t(t,r) \neq 0$, $W_r(t,r) \neq 0$, $W_t(t,r)W_r(t,r) + 2W_{tr}(t,r) = 0$, $W_r^2(t,r) + 2W_{rr}(t,r) = 0$ and $W_t^2(t,r) + 2W_{tt}(t,r) = 0$.

(D13) Rank=1, $U_t(t,r) \neq 0$, $U_r(t,r) = 0$, $V_t(t,r) = 0$, $V_r(t,r) = 0$, $W_t(t,r) \neq 0$, $W_r(t,r) \neq 0$, $W_t(t,r)W_r(t,r) + 2W_{tr}(t,r) = 0$ and $W_r^2(t,r) + 2W_{rr}(t,r) = 0$.

(D14) Rank=1, $U_t(t,r) = 0$, $U_r(t,r) = 0$, $V_t(t,r) \neq 0$, $V_r(t,r) \neq 0$, $W_t(t,r) = 0$, $W_r(t,r) = 0$, $V_t(t,r)V_r(t,r) + 2V_{tr}(t,r) = 0$, $V_t^2(t,r) + 2V_{tt}(t,r) = 0$ and $V_r^2(t,r) + 2V_{rr}(t,r) \neq 0$.

(D15) Rank=1, $U_t(t,r) = 0$, $U_r(t,r) = 0$, $V_t(t,r) = 0$, $V_r(t,r) = 0$, $W_t(t,r) \neq 0$, $W_r(t,r) \neq 0$, $W_t(t,r)W_r(t,r) + 2W_{tr}(t,r) = 0$, $W_t^2(t,r) + 2W_{tt}(t,r) = 0$ and $W_r^2(t,r) + 2W_{rr}(t,r) \neq 0$.

(D16) Rank=1, $U_t(t,r) = 0$, $U_r(t,r) = 0$, $V_t(t,r) \neq 0$, $V_r(t,r) = 0$, $W_t(t,r) \neq 0$, $W_r(t,r) = 0$, $W_t^2(t,r) + 2W_{tt}(t,r) = 0$ and $V_t^2(t,r) + 2V_{tt}(t,r) = 0$.

(D17) Rank=1, $U_t(t,r) \neq 0$, $U_r(t,r) = 0$, $V_t(t,r) \neq 0$, $V_r(t,r) = 0$, $W_t(t,r) = 0$, $W_r(t,r) \neq 0$, $W_r^2(t,r) + 2W_{rr}(t,r) = 0$ and $V_t^2(t,r) + 2V_{tt}(t,r) = 0$.

(D18) Rank=1, $U_t(t,r) = 0$, $U_r(t,r) = 0$, $V_t(t,r) \neq 0$, $V_r(t,r) = 0$, $W_t(t,r) = 0$, $W_r(t,r) \neq 0$, $W_r^2(t,r) + 2W_{rr}(t,r) \neq 0$ and $V_t^2(t,r) + 2V_{tt}(t,r) = 0$.

(D19) Rank=1, $U_t(t,r) = 0$, $U_r(t,r) = 0$, $V_t(t,r) \neq 0$, $V_r(t,r) \neq 0$, $W_t(t,r) = 0$, $W_r(t,r) \neq 0$, $V_t(t,r)V_r(t,r) + 2V_{tr}(t,r) = 0$, $V_t^2(t,r) + 2V_{tt}(t,r) = 0$, $V_r^2(t,r) + 2V_{rr}(t,r) = 0$ and $W_r^2(t,r) + 2W_{rr}(t,r) = 0$.

(D20) Rank=1, $U_t(t,r) = 0$, $U_r(t,r) = 0$, $V_t(t,r) \neq 0$, $V_r(t,r) \neq 0$, $W_t(t,r) \neq 0$, $W_r(t,r) = 0$, $V_t(t,r)V_r(t,r) + 2V_{tr}(t,r) = 0$, $V_t^2(t,r) + 2V_{tt}(t,r) = 0$, $V_r^2(t,r) + 2V_{rr}(t,r) = 0$ and $W_t^2(t,r) + 2W_{tt}(t,r) = 0$.

(D21) Rank=1, $U_t(t,r) = 0$, $U_r(t,r) = 0$, $V_t(t,r) \neq 0$, $V_r(t,r) = 0$, $W_t(t,r) = 0$, $W_r(t,r) \neq 0$, $V_t^2(t,r) + 2V_{tt}(t,r) \neq 0$ and $W_r^2(t,r) + 2W_{rr}(t,r) = 0$.

We will consider each case in turn.

## Case A1

In this case we have $U_t(t,r) = 0$, $U_r(t,r) = 0$, $V_t(t,r) = 0$, $V_r(t,r) \neq 0$, $V_r^2(t,r) + 2V_{rr}(t,r) = 0$, $W_t(t,r) \neq 0$, $W_r(t,r) \neq 0$, $W_t(t,r)W_r(t,r) + 2W_{tr}(t,r) \neq 0$, $W_t^2(t,r) + 2W_{tt}(t,r) \neq 0$, $W_r^2(t,r) + 2W_{rr}(t,r) \neq 0$ and the rank of the $6 \times 6$ Riemann matrix is 3 and there exists no non trivial solution of equation (3). From the above constraints we get $U(t,r) = a$, $V(t,r) = \ln(br+c)^2$, $W = W(t,r)$, where $a,b,c \in R(b \neq 0)$. Substituting the information of $U(t,r)$ and $V(t,r)$ in (7) and after a rescaling of $t$, the line element can be written in the form

$$ds^2 = -dt^2 + dr^2 + (br+c)^2 d\theta^2 + e^{W(t,r)} d\phi^2. \qquad (10)$$

This case belongs to the class A. In class A the rank of the $6 \times 6$ Riemann matrix is 6, 5, 4, 3 or 2 (excluding the class B) and there exists no non trivial solution of equation (3) [7]. The class A is said to generic in the sense that every curvature symmetry is necessarily a homothetic vector field [7,16]. Hence in this case no proper curvature symmetry exists. Cases (A2) to (A30) are precisely the same.

## Case B1

In this case we have $U_t(t,r) \neq 0$, $U_r(t,r) = 0$, $V_t(t,r) \neq 0$, $V_r(t,r) = 0$, $V_t^2(t,r) + 2V_{tt}(t,r) = 0$, $W_r(t,r) \neq 0$, $W_r^2(t,r) + 2W_{rr}(t,r) \neq 0$, $W_t(t,r) = 0$ and the rank of the $6 \times 6$ Riemann matrix is two. There exists no solution of equation (3). From the above constraints we have $U = U(t)$, $V = \ln(at+b)$ and $W = W(r)$, where $a, b \in R (a \neq 0)$. Substituting the information of $U(t,r)$, $V(t,r)$ and $W(t,r)$ in (7), the line element can be written in the form

$$ds^2 = -e^{U(t)} dt^2 + (at+b)^2 d\theta^2 + dr^2 + e^{W(r)} d\phi^2. \tag{11}$$

The above space-time is clearly 2+2 decomposable and belongs to curvature class B. CCS in this case take the form [7]

$$X = X_1 + X_2, \tag{12}$$

where $X_1$ are CCS in the induced geometry on each of the two dimensional submanifolds of constant $r$ and $\phi$, and $X_2$ are CCS in the induced geometry on each of the two dimensional submanifolds of constant $t$ and $\theta$. The next step is to work out the CCS in the induced geometry of the submanifolds of constant $r$ and $\phi$. A method for finding CCS in 2-dimensional submanifolds is given in [16]. If one proceeds further the given nonzero components of the induced metric on each of the two dimensional submanifolds of constant $r$ and $\phi$ are given by

$$g_{00} = -e^U, \qquad g_{22} = (at+b)^2 \tag{13}$$

and the Ricci scalar is $R = -\frac{1}{2} \dot{U} a e^{-U}$ then there exist following two possibilities:

$$(\alpha)\ E = q \qquad (\beta)\ E \neq q,$$

where

$$E \equiv \left( \frac{\left( (at+b) \dot{U} e^{-U} \right) e^U}{2(at+b)^{\frac{1}{2}} \dot{U}^{\frac{3}{2}}} \right)^{\cdot} (at+b)^{\frac{3}{2}} \dot{U}^{\frac{1}{2}} e^{-U}$$

and $q \in R$. First consider the case ($\alpha$), in which further three more possibilities exist which are:

(i) $q > 0$, (ii) $q < 0$, (iii) $q = 0$.

We will consider each possibility in turn.

($\alpha i$) CCS in this case are

$$X^0 = \left(c_2 \cosh \sqrt{q}\,\theta + c_3 \sinh \sqrt{q}\,\theta\right)\left(\frac{\dot{U}}{at+b}\right)^{-\frac{1}{2}},$$

$$X^2 = \sqrt{q}\left(c_3 \cosh \sqrt{q}\,\theta + c_2 \sinh \sqrt{q}\,\theta\right)\int (at+b)^{-\frac{3}{2}} \dot{U}^{-\frac{1}{2}} e^U\, dt + c_4 \tag{14}$$

provided that $E = q$, where $c_2, c_3, c_4 \in R$.

($\alpha\,ii$) In this case $q < 0$. Put $q = -Q$, where $Q \in R(Q > 0)$. CCS in this case

$$X^0 = \left(c_2 \cos \sqrt{Q}\,\theta + c_3 \sin \sqrt{Q}\,\theta\right)\left(\frac{\dot{U}}{at+b}\right)^{-\frac{1}{2}},$$

$$X^2 = \sqrt{Q}\left(c_3 \cos \sqrt{Q}\,\theta - c_2 \sin \sqrt{Q}\,\theta\right)\int (at+b)^{-\frac{3}{2}} \dot{U}^{-\frac{1}{2}} e^U\, dt + c_4, \tag{15}$$

provided that $E = -Q$, where $c_2, c_3, c_4 \in R$.

($\alpha\,iii$) In this case $q = 0$. CCS in this case

$$X^0 = (c_2 + c_3\theta)\left(\frac{\dot{U}}{at+b}\right)^{-\frac{1}{2}},$$

$$X^2 = c_3 \int \left(\frac{\dot{U}}{at+b}\right)^{-\frac{1}{2}} \frac{e^U}{(at+b)^2}\, dt - N\left(\frac{c_3}{2}\theta^2 + c_2\theta\right) + c_4, \tag{16}$$

provided that

$$\frac{\left((at+b)\dot{U}\,e^{-U}\right)\dot{}\,e^U}{2(at+b)^{\frac{1}{2}}\dot{U}^{\frac{3}{2}}} = N,$$

where $c_2, c_3, c_4, N \in R(N \neq 0)$.

($\beta$) CCS in this case are

$$X^0 = 0, \quad X^2 = c_5, \tag{17}$$

where $c_5 \in R$.

Now we are interested to work out the CCS in the induced geometry of the submanifolds of constant $t$ and $\theta$. The induced metric on each of the two dimensional submanifolds of constant $t$ and $\theta$ are given by

$$g_{11} = 1, \qquad g_{33} = e^{W(r)} \tag{18}$$

and the Ricci scalar is $R = -\frac{1}{2}(W'^2(r) + 2W''(r))$ then there exist following two possibilities:

($\lambda$) $O = m$         ($\delta$) $O \neq m$,

where

$$O = \left(\frac{\left((W'^2(r) + 2W''(r))e^{W(r)}\right)'}{2(W'^2(r) + 2W''(r))^{\frac{3}{2}}} e^{-W(r)}\right)' (W'^2(r) + 2W''(r))^{\frac{1}{2}} e^{W(r)}$$

and $m \in R$. First consider the case ($\lambda$), in which further three more possibilities exist which are:

(i) $m > 0$,     (ii) $m < 0$,     (iii) $m = 0$.

We will consider each possibility in turn.

($\lambda$ i)    CCS in this case are

$$\begin{aligned} X^1 &= \left(c_6 \cosh\sqrt{m}\,\phi + c_7 \sinh\sqrt{m}\,\phi\right)(W'^2(r) + 2W''(r))^{-\frac{1}{2}}, \\ X^3 &= \sqrt{m}\left(c_7 \cosh\sqrt{m}\,\phi + c_6 \sinh\sqrt{m}\,\phi\right)\int (W'^2(r) + 2W''(r))^{-\frac{1}{2}} e^{-W(r)} dr + c_8, \end{aligned} \qquad (19)$$

provided that $O = m$, where $c_6, c_7, c_8 \in R$.

($\lambda$ ii)    In this case $m < 0$. Put $m = -Q_1$, where $Q_1 \in R (Q_1 > 0)$. CCS in this case

$$\begin{aligned} X^1 &= \left(c_6 \cos\sqrt{Q_1}\,\phi + c_7 \sin\sqrt{Q_1}\,\phi\right)(W'^2(r) + 2W''(r))^{-\frac{1}{2}}, \\ X^3 &= \sqrt{Q_1}\left(c_7 \cos\sqrt{Q_1}\,\phi - c_6 \sin\sqrt{Q_1}\,\phi\right)\int (W'^2(r) + 2W''(r))^{-\frac{1}{2}} e^{-W(r)} dr + c_8, \end{aligned} \qquad (20)$$

provided that $O = -Q_1$, where $c_6, c_7, c_8 \in R$.

($\lambda$ iii)    In this case $m = 0$. CCS in this case

$$\begin{aligned} X^1 &= (c_6 \phi + c_7)(W'^2 + 2W'')^{-\frac{1}{2}}, \\ X^3 &= c_6 \int (W'^2 + 2W'')^{-\frac{1}{2}} e^{-W} dr - N_1 \left(\frac{c_6}{2}\phi^2 + c_7 \phi\right) + c_8, \end{aligned} \qquad (21)$$

provided that

$$\frac{((W'^2 + 2W'')e^W)'}{2(W'^2 + 2W'')^{\frac{3}{2}}} e^{-W} = N_1,$$

where $c_6, c_7, c_8, N_1 \in R (N_1 \neq 0)$.

($\delta$)    CCS in this case are

$$X^1 = 0, \quad X^3 = c_6, \tag{22}$$

where $c_6 \in R$. CCS in this case are given in equation (12), either of (14), (15), (16) or (17) along with (19), (20), (21) or (22). This has been omitted here for the sake of brevity. Cases (B2) and (B3) are precisely the same.

**Case C1**

In this case we $U_t(t,r) \neq 0$, $U_r(t,r) = 0$, $V_t(t,r) = 0$, $V_r(t,r) \neq 0$, $V_r^2(t,r) + 2V_{rr}(t,r) \neq 0$, $W_t(t,r) = 0$, $W_r(t,r) \neq 0$, $W_r^2(t,r) + 2W_{rr}(t,r) \neq 0$ and the rank of the $6 \times 6$ Riemann matrix is three and there exists a unique (up to multiple) no where zero vector field $t_a = t_{,a}$ such that $t_{a;b} = 0$. From the Ricci identity $R_{abcd} t^d = 0$. The above constraints give $U = U(t)$, $V = V(r)$ and $W = W(r)$. After a suitable rescaling of $t$, the line element can be written in the form

$$ds^2 = -dt^2 + dr^2 + e^{V(r)} d\theta^2 + e^{W(r)} d\phi^2. \tag{23}$$

The above space-time is clearly 1+3 decomposable and belongs to curvature class C. CCS in this case [7] are

$$X = f(t)\frac{\partial}{\partial t} + X', \tag{24}$$

where $f(t)$ is an arbitrary function of $t$ and $X'$ is a homothetic vector field in the induced geometry on each of the three-dimensional submanifolds of constant $t$. The induced metric $g_{\alpha\beta}$ (where $\alpha, \beta = 1, 2, 3$) with non zero components is given by

$$g_{11} = 1, \qquad g_{22} = e^V, \qquad g_{22} = e^W. \tag{25}$$

A vector field $X'$ is called homothetic vector field if it satisfies

$$L_{X'} g_{\alpha\beta} = 2c\, g_{\alpha\beta}, \qquad c \in R. \tag{26}$$

One can expand the above equation (26) using (25) to get

$$X^1{}_{,1} = c, \tag{27}$$

$$X^1{}_{,2} + e^V X^2{}_{,1} = 0, \tag{28}$$

$$X^1{}_{,3} + e^W X^3{}_{,1} = 0, \tag{29}$$

$$V' X^1 + 2 X^2{}_{,2} = 2c, \tag{30}$$

$$e^V X^2{}_{,3} + e^W X^3{}_{,2} = 0, \tag{31}$$

$$W' X^1 + 2 X^3{}_{,3} = 2c. \tag{32}$$

Solving equations (27), (28) and (29), we get

$$X^1 = cr + A^1(\theta,\phi), \quad X^2 = -A^1{}_\theta(\theta,\phi)\int e^{-V} dr + A^2(\theta,\phi),$$
$$X^3 = -A^1{}_\phi(\theta,\phi)\int e^{-W} dr + A^3(\theta,\phi), \tag{33}$$

where $A^1(\theta,\phi)$, $A^2(\theta,\phi)$ and $A^3(\theta,\phi)$ are functions of integration. If one proceeds further one finds that there arise two possibilities which are

(i) $(V - W)' \neq 0$ \qquad (ii) $(V - W)' = 0$.

In sub case (i) the solution of the above equations from (27) to (32) is

$$X^1 = cr + c_1, \quad X^2 = \theta c_2 + c_3, \quad X^3 = \phi c_4 + c_5, \tag{34}$$

where $\quad V(r) = \ln(cr + c_1)^{2(1-\frac{c_2}{c})}, \quad W(r) = \ln(cr + c_1)^{2(1-\frac{c_4}{c})} \quad$ and $c, c_1, c_2, c_3, c_4, c_5 \in R(c_2 \neq c_4, c_2 \neq 0, c_4 \neq 0, c \neq 0, c_2 \neq c, c_4 \neq c)$. It follows from the above calculation that the induced metric in the induced geometry on each of the three-dimensional submanifolds of constant $t$ admits proper homothetic vector field. CCS in this case are given by the use of equations (24) and (34) as

$$X^0 = f(t), \quad X^1 = cr + c_1, \quad X^2 = \theta c_2 + c_3, \quad X^3 = \phi c_4 + c_5, \tag{35}$$

where $f(t)$ is an arbitrary function of $t$. One can write the above equation (35) after subtracting homothetic vector fields as

$$X = (f(t), 0, 0, 0). \tag{36}$$

Clearly CCS form an infinite dimensional vector space.

In sub case (ii) we have $(V - W)' = 0$ $V = W + \lambda$, where $\lambda \in R$ and $V'' \neq 0$ (and so $W'' \neq 0$). Solution of the above equations from (27) to (32) is

$$X^1 = cr + c_1, \quad X^2 = \theta c_2 + \phi c_3 + c_4, \quad X^3 = \phi c_2 - \theta e^\lambda c_3 + c_5, \tag{37}$$

where $V(r) = \ln(cr + c_1)^{2(1-\frac{c_2}{c})} + \lambda$ and $c, c_1, c_2, c_3, c_4, c_5 \in R(c_2 \neq c, c \neq 0, c_2 \neq 0)$. It follows from the above calculation that the induced metric in the induced geometry on each of the three-dimensional submanifolds of constant $t$ admits proper homothetic vector field. CCS in this case are given by the use of equations (24) and (37) as

$$X^0 = f(t), \quad X^1 = cr + c_1, \quad X^2 = \theta c_2 + \phi c_3 + c_4, \quad X^3 = \phi c_2 - \theta e^\lambda c_3 + c_5. \tag{38}$$

One can write the above equation (38) after subtracting homothetic vector fields one has equation (36). CCS clearly form an infinite dimensional vector space. Cases C2 to C42 are precisely the same.

**Case C43**

In this case we have $U_t(t,r) = 0$, $U_r(t,r) = 0$, $V_t(t,r) \neq 0$, $V_r(t,r) \neq 0$, $V_t^2(t,r) + 2V_{tt}(t,r) = 0$, $V_t(t,r)V_r(t,r) + 2V_{tr}(t,r) = 0$, $V_r^2(t,r) + 2V_{rr}(t,r) \neq 0$, $W_t(t,r) \neq 0$, $W_t^2(t,r) + 2W_{tt}(t,r) = 0$, $W_t(t,r)W_r(t,r) + 2W_{tr}(t,r) = 0$, $W_r(t,r) \neq 0$, $W_r^2(t,r) + 2W_{rr}(t,r) \neq 0$ and the rank of the $6 \times 6$ Riemann matrix is three and there exists a unique (up to multiple) $t_a = t_{,a}$ solution of equation (3) but $t_a$ is not covariantly constant. From the above constraints we have $U = k$, $V(t,r) = \ln(at + b + \alpha(r))^2$ and $W(t,r) = \ln(ct + d + \beta(r))^2$, where $a, b, c, d, k \in R(a, c \neq 0)$ and $\alpha(r)$ and $\beta(r)$ are no where zero and no where equal functions of integration. After a suitable rescaling of $t$, the line element can be written in the form

$$ds^2 = -dt^2 + dr^2 + (at + b + \alpha(r))^2 d\theta^2 + (ct + d + \beta(r))^2 d\phi^2. \qquad (39)$$

Substituting the above information into CC equations and after simplifying we have

$$X^0_{,1} = X^0_{,2} = X^0_{,3} = 0, \quad X^1 = 0, \quad X^2_{,0} = X^2_{,1} = X^2_{,2} = X^2_{,3} = 0,$$
$$X^3_{,0} = X^3_{,1} = X^3_{,2} = X^3_{,3} = 0.$$

From the above equations one has $X^0 = f(t)$, $X^1 = 0$, $X^2 = c_1$ and $X^3 = c_2$, where $c_1, c_2 \in R$. CCS in this case

$$X^0 = f(t), \quad X^1 = 0, \quad X^2 = c_1, \quad X^3 = c_2, \qquad (40)$$

where $f(t)$ is an arbitrary function. One can write the above equation (40) after subtracting Killing vector fields one gets equation (36). CCS clearly form an infinite dimensional vector space. Cases C44 to C64 are precisely the same.

**Case D1**

In this we have $U_t(t,r) = 0$, $U_r(t,r) = 0$, $V_t(t,r) = 0$, $V_r(t,r) = 0$, $W_t(t,r) = 0$, $W_r(t,r) \neq 0$, $W_r^2(t,r) + 2W_{rr}(t,r) \neq 0$ and the rank of the $6 \times 6$ Riemann matrix is 1. Here there exist two linearly independent solutions $t_a = t_{,a}$ and $\theta_a = \theta_{,a}$ of equation (3) and satisfying $t_{a;b} = 0$ and $\theta_{a;b} = 0$. From the above constraints we have $U = k_1$, $V = k_2$ and $W = W(r)$, where $k_1, k_2 \in R$. The line element, after rescaling of $t$ and $\theta$, can be written as

$$ds^2 = -dt^2 + dr^2 + d\theta^2 + e^{W(r)}d\phi^2. \tag{41}$$

The above space-time (56) is clearly 1+1+2 decomposable and belongs to the curvature class D. CCS in this case are [7]

$$X = f(t,\theta)\frac{\partial}{\partial t} + g(t,\theta)\frac{\partial}{\partial \theta} + X', \tag{42}$$

where $f(t,\theta)$ and $g(t,\theta)$ are arbitrary functions of $t$ and $\theta$ and $X'$ is a CC on each of two dimensional submanifolds of constant $t$ and $\theta$. CCS on each of two dimensional submanifolds of constant $t$ and $\theta$ are given in case B1. This has been omitted here for the sake of brevity. Clearly CCS in this case form an infinite dimensional vector space. Cases (D2) to (D8) are precisely the same.

**Case D9**

In this case we have $U_t(t,r) \neq 0$, $U_r(t,r) = 0$, $V_t(t,r) \neq 0$, $V_r(t,r) \neq 0$, $V_t(t,r)V_r(t,r) + 2V_{tr}(t,r) = 0$, $V_r^2(t,r) + 2V_{rr}(t,r) = 0$, $W_t(t,r) = 0$, $W_r(t,r) = 0$ and the rank of the $6 \times 6$ Riemann matrix is 1. From the above constraints we have $U = U(t)$, $V = \ln(ar+b+\sigma(t))^2$ and $W = k_1$, where $a,b,k_1 \in R(a \neq 0)$ and $\sigma(t)$ is nowhere zero function of integration. Here there exist two linearly independent solutions $r_a = r_{,a}$ and $\phi_a = \phi_{,a}$ of equation (3). The vector field $r_a$ is not covariantly constant where as $\phi_a$ is covariantly constant. The line element, after rescaling of $\phi$, can be written as

$$ds^2 = -e^{U(t)}dt^2 + dr^2 + (ar+b+\sigma(t))^2 d\theta^2 + d\phi^2. \tag{43}$$

The above space-time (43) is clearly 1+3 decomposable and belongs to the curvature class D. Substituting the above information in CC equations one finds that CCS in this case are

$$X^0 = X^1 = 0, \quad X^2 = c_1, \quad X^3 = f(r,\phi), \tag{44}$$

where $f(r,\phi)$ is an arbitrary function and $c_1 \in R$. One can write the above equation (44) after subtracting Killing vector fields as

$$X = (0,0,0,f(r,\phi)). \tag{45}$$

Clearly CCS form an infinite dimensional vector space. Cases (D9) to (D18) are precisely the same.

**Case D19**

In this case we have the conditions $U_t(t,r) = 0$, $U_r(t,r) = 0$, $V_t(t,r) \neq 0$, $V_r(t,r) \neq 0$, $V_t(t,r)V_r(t,r) + 2V_{tr}(t,r) = 0$, $V_r^2(t,r) + 2V_{rr}(t,r) = 0$, $W_t(t,r) = 0$, $V_t^2(t,r) + 2V_{tt}(t,r) = 0$, $W_r(t,r) \neq 0$, $W_r^2(t,r) + 2W_{rr}(t,r) = 0$ and the rank of the $6 \times 6$ Riemann matrix is 1. From the above constraints we have $U = k_1$, $V = \ln(at+br+c)^2$ and $W = \ln(pr+q)^2$, where $a,b,c,p,q,k_1 \in R (a,b,p \neq 0)$. In this case there exist two linearly independent solutions $t_a = t_{,a}$ and $r_a = r_{,a}$ of equation (3). The vector fields $t_a$ and $r_a$ are not covariantly constant and the line element, after rescaling of $t$, can be written as

$$ds^2 = -dt^2 + dr^2 + (at+br+c)^2 d\theta^2 + (pr+q)^2 d\phi^2. \tag{46}$$

The above space-time (46) belongs to the curvature class D. Substituting the above information in CC equations one finds CCS in this case are

$$X^0 = M(t,r), \qquad X^1 = 0, \qquad X^2 = c_1, \qquad X^3 = c_2, \tag{47}$$

where $M(t,r)$ is an arbitrary function and $c_1, c_2 \in R$. One can write the above equation (47) after subtracting Killing vector fields as

$$X = (M(t,r), 0, 0, 0). \tag{48}$$

Clearly, CCS form an infinite dimensional vector space. Cases (D20) and (D21) are precisely the same.

## SUMMARY

In this paper a study of non-static cylindrically symmetric space-times according to their proper curvature symmetries is given. An approach is adopted to study the above space-times by using the rank of the $6 \times 6$ Riemann matrix and also using the theorem given in [7], which suggested where proper curvature symmetries exist. From the above study we obtain the following results:

(i) The case when the rank of the $6 \times 6$ Riemann matrix is three and there exists a unique no where zero independent timelike vector field which is a solution of equation (3) and is covariantly constant. This is the space-time (23) and it admits proper CCS which form an infinite dimensional vector space (see case C1).

(ii) The case when the rank of the $6 \times 6$ Riemann matrix is three and there exists a unique nowhere zero independent timelike vector field which is a solution of equation (3) and is not covariantly constant. This is the space-time (39) and it admits proper CCS which form an infinite dimensional vector space (see case C43).

(iii) The case when the rank of the $6 \times 6$ Riemann matrix is one and there exist two nowhere zero independent vector fields which are solutions of equation (3) and are covariantly constant. This is the space-time (41) and it admits proper CCS, which form an infinite dimensional vector space (see case D1).

(iv)    The case when the rank of the $6\times 6$ Riemann matrix is one and there exist two nowhere zero independent solutions of equation (3) but only one independent covariantly constant vector field. This is the space-time (43) and it admits proper CCS which form an infinite dimensional vector space (see case D9).

(v)     The case when the rank of the $6\times 6$ Riemann matrix is one and there exist two nowhere zero independent solutions of equation (3) but are not covariantly constant vector field. This is the space-time (46) and it admits proper CCS which form an infinite dimensional vector space (see case D19).

(vi)    The case when the rank of the $6\times 6$ Riemann matrix is two and there exists no non trivial solution of equation (3). This is the space-times (11) and it admits CCS which are Killing vector fields (see case B1).